\documentclass[10pt,journal,compsoc,onecolumn]{IEEEtran}
\usepackage{cite}
\usepackage[pdftex]{graphicx}
\graphicspath{{../pdf/}{../jpeg/}}
\DeclareGraphicsExtensions{.pdf,.jpeg,.png}
\usepackage{array}
\usepackage{amsmath}
\usepackage{amssymb}
\usepackage{mdwmath}
\usepackage{mdwtab}
\usepackage{eqparbox}
\usepackage{url}
\usepackage{hyperref}
\usepackage[switch]{lineno}
\usepackage{xcolor}

\usepackage{tcolorbox}
\tcbset{standard jigsaw, opacityback=0.5}

\begin{document}
% \linenumbers
\title{Unleashing the Strengths of Unlabeled Data in Pan-cancer Abdominal Organ Quantification: \\ the FLARE22 Challenge}

\author{Jun Ma, Yao Zhang, Song Gu, Cheng Ge, Shihao Ma, Adamo Young,  Cheng Zhu, Kangkang Meng, Xin Yang, Ziyan Huang, Fan Zhang, Wentao Liu, YuanKe Pan, Shoujin Huang, Jiacheng Wang, Mingze Sun, Weixin Xu, Dengqiang Jia, Jae Won Choi, Nat\'{a}lia Alves, Bram de Wilde, Gregor Koehler, Yajun Wu, Manuel Wiesenfarth, Qiongjie Zhu, Guoqiang Dong, Jian He, the FLARE Challenge Consortium, and\\ Bo Wang % <-this % stops a space
\IEEEcompsocitemizethanks{
\IEEEcompsocthanksitem A full list of affiliations appears at the end of the paper. Corresponding author: Bo Wang.
E-mail: bowang@vectorinstitute.ai}% <-this % stops an unwanted space
% \thanks{Manuscript received Jan. 10, 2023}
}

\IEEEtitleabstractindextext{%
\begin{abstract}
Quantitative organ assessment is an essential step in automated abdominal disease diagnosis and treatment planning. Artificial intelligence (AI) has shown great potential to automatize this process. 
However, most existing AI algorithms rely on many expert annotations and lack a comprehensive evaluation of accuracy and efficiency in real-world multinational settings. To overcome these limitations, we organized the FLARE 2022 Challenge, the largest abdominal organ analysis challenge to date, to benchmark fast, low-resource, accurate, annotation-efficient, and generalized AI algorithms.
We constructed an intercontinental and multinational dataset from more than 50 medical groups, including Computed Tomography (CT) scans with different races, diseases, phases, and manufacturers. 
We independently validated that a set of AI algorithms achieved a median Dice Similarity Coefficient (DSC) of 90.0\% by using 50 labeled scans and 2000 unlabeled scans, which can significantly reduce annotation requirements. 
% The winning algorithm can precisely annotate 13 organs for a 3D CT scan with 50 million voxels using no more than 10s, 2GB GPU memory, and 25\% CPU utilization on average, which can be deployed in low-resource settings.
The best-performing algorithms successfully generalized to holdout external validation sets, achieving a median DSC of 89.5\%, 90.9\%, and 88.3\% on North American, European, and Asian cohorts, respectively. They also enabled automatic extraction of key organ biology features, which was labor-intensive with traditional manual measurements.
This opens the potential to use unlabeled data to boost performance and alleviate annotation shortages for modern AI models.
\end{abstract}

}

\maketitle

\IEEEdisplaynontitleabstractindextext

\IEEEpeerreviewmaketitle

\section*{Introduction}\label{intro}
% P1. abdomen cancer is very common; the clinical practice requires automatic tools
Abdominal organs are high cancer incidence areas, such as liver cancer, kidney cancer, pancreas cancer, and gastric cancer~\cite{CancerStatis22}. Computed Tomography (CT) scanning has been a major imaging technology for the diagnosis and treatment of abdominal cancer because it can yield important prognostic information with fast imaging speed for cancer patients, which has been recommended by many clinical treatment guidelines.
In order to quantify abdominal organs, radiologists and clinicians need to manually delineate organ boundaries in each slice of the 3D CT scans~\cite{Summers21-LiverVol, SpleenVol}. 
However, manual segmentation is time-consuming and inherently subjective with inter- and intra-expert variability. Inaccurate segmentation and quantification can lead to over-dosing or under-dosing treatment, increasing the risk of injuries.
Therefore, accurate and automatic segmentation and quantification tools are highly demanded for abdominal cancer treatment in clinical practice.

% P2. current progress 
AI has shown great promise for automatic cancer quantification and diagnosis in medical images~\cite{NatMed-Retina2018, Nature-HeartAI, NatMed-Heart, NatMed21-Prostate}. In the field of abdominal cancer CT analysis, a large number of automatic segmentation algorithms have been proposed during the past decade, specifically for liver cancer~\cite{H-DenseUNet}, kidney cancer~\cite{KiTS2021MIA}, and pancreas cancer~\cite{MICCAI15-Pancreas}. However, these algorithms face limitations due to their reliance on costly annotations and lack of comprehensive evaluation on multinational datasets. AI competitions have been an effective way to gather efforts from the whole research community to solve hard tasks and promote methodology developments~\cite{NatMeth-nucleus, MSD-Summary, NatMed-panda}. Several abdominal cancer-related AI challenges have been organized~\cite{heimann2009Sliver, LiTS-MIA, KiTS-Challenge}. These challenges greatly facilitated innovations in liver and kidney cancer image analysis but the algorithms were not evaluated on external datasets and the evaluation metrics only focused on segmentation accuracy without considering algorithm efficiency.
Therefore, there is an unmet need for annotation-efficient and universal abdominal organ segmentation algorithms that are robust to different diseases, races, and imaging platforms.

To address the above limitations, we organized a global AI challenge, the Fast and Low-resource semi-supervised Abdominal oRgan sEgmentation (FLARE), with diverse abdominal CT images and novel challenge design to prompt the development of annotation- and resource-efficient AI algorithms for abdominal organ quantification. 
Specifically, we constructed a large-scale multi-racial, multi-center, multi-disease, multi-phase, and multi-manufacturer abdominal CT dataset from 2900 patients, including 725,000 slices, 53 medical groups, seven CT scanner manufacturers, and four CT phases and covering more than six types of abdominal cancer. This is the largest and most diverse publicly available dataset to date. 
Furthermore, we designed a semi-supervised competition task, where participants were provided with a limited set of annotated images and a large number of unlabeled images. The evaluation metrics focus on both segmentation accuracy and efficiency. In order to validate generalization ability, top-performing algorithms were externally evaluated on independent Asian, European, and North American cohorts. 
The validation datasets were blind to participants, minimizing the risk of data leakage and providing a fair platform for benchmarking. 
This is the first time that AI algorithms for abdominal organ quantification have been challenged to learn with limited annotations and generalize on unseen medical centers without user interaction or additional fine-tuning. The dataset, top algorithms, and benchmark platform have been made publicly available for reproduction and long-term algorithm comparison.

\begin{figure*}[!htbp]%
\centering
\includegraphics[scale=0.4]{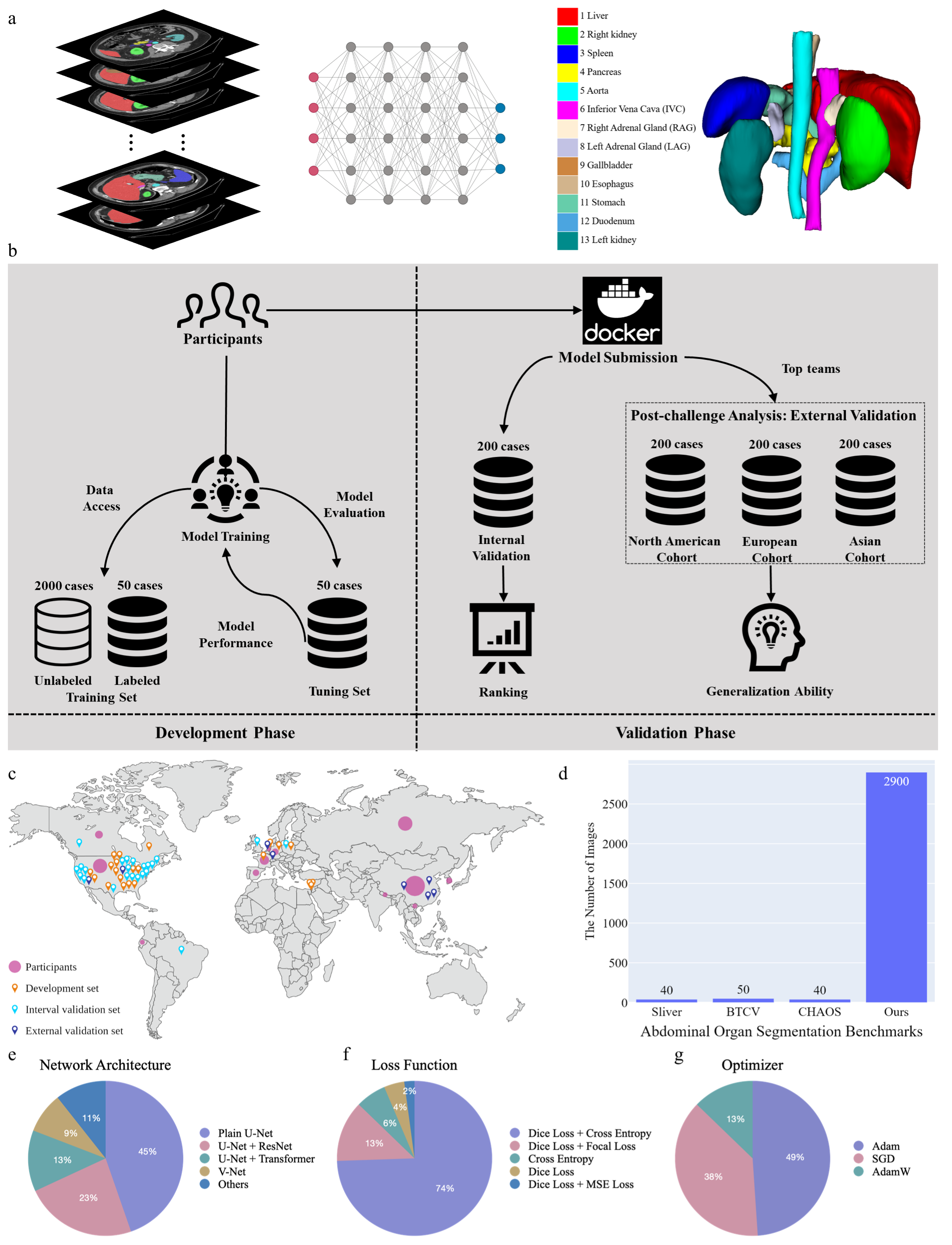}
\caption{\textbf{Overview of the challenge design.} \textbf{a,} The challenge aims to benchmark automatic algorithms that can simultaneously segment 13 abdominal organs. Organs have different sizes, morphologies, and appearances, featuring representative difficulties in medical image analysis tasks. \textbf{b,} The challenge contains two phases. During the development phase, participants develop automatic segmentation algorithms based on 2000 unlabeled cases and 50 labeled cases. The algorithms can be evaluated on the tuning set and the online evaluation platform will return the quantitative performance to participants. During the validation phase, each participant team can submit one algorithm via the docker container as the final solution, which is independently evaluated on the internal validation set to obtain ranking results. The top teams are selected for post-challenge analysis, which are further evaluated on three independent intercontinental cohorts to validate their generalization ability. 
\textbf{c,} The data sources are multinational and the challenge has attracted more than 100 worldwide participants (the circle size is proportional to the number of participants in each country). \textbf{d,} The FLARE challenge dataset is significantly larger than the previous abdominal organ segmentation challenge datasets. Distribution of key algorithm designs: \textbf{e} network architecture, \textbf{f,} loss function, and \textbf{g,} optimizer. 
}\label{fig:overview}
\end{figure*}

\section*{Results}\label{sec2}
\subsection*{Challenge design}
The FLARE 2022 challenge followed two guidelines: Enhancing the QUAlity and Transparency Of health Research (EQUATOR, \url{https://www.equator-network.org/}) and Biomedical Image Analysis ChallengeS (BIAS)~\cite{maier2020BIAS-MIA}, which has been preregistered and passed peer review (two technical reviews and one clinical review)~\cite{MA-FLARE22-Design}. 
The challenge aims to benchmark algorithms that can automatically segment 13 abdominal organs from CT scans, including the liver, right kidney, spleen, pancreas, aorta,  inferior vena cava, right adrenal gland, left adrenal gland, gallbladder, esophagus, stomach, duodenum, and left kidney (Fig.~\ref{fig:overview}a). These organs were selected based on clinical needs and featured representative difficulties that are usually encountered in medical image analysis tasks.

Motivated by the real-world setting, we designed a semi-supervised segmentation task (Fig.~\ref{fig:overview}b) rather than the typical fully supervised learning task because an abundance of unlabeled medical images exists in most medical centers, and collecting a small annotated dataset is feasible. Specifically, participants were provided with 2000 unlabeled cases and 50 labeled cases to develop automatic segmentation algorithms for abdominal organ segmentation.  
The challenge contained two phases. During the development phase, participants can access the whole training set and the 50 images in the tuning set. The tuning set labels were deployed on the online evaluation platform that was not publicly available, but participants can submit their results to the platform and get feedback on the algorithm performance. 

During the validation phase, each team had one chance to submit one algorithm as the final solution. Different from most of the existing challenges\cite{NatMed-panda, NatMeth-nucleus} that only considered accuracy-related metrics during ranking, our evaluation metrics focused on both segmentation accuracy and efficiency. In particular, we evaluated the degree of region overlap (Dice Similarity Coefficient, DSC), the degree of boundary matching (Normalized Surface Distance, NSD), running time, GPU memory consumption, and CPU utilization (Methods). This is the first time that medical image segmentation algorithms have been quantitatively evaluated with both accuracy-related metrics and comprehensive resource (i.e., GPU and CPU) consumption metrics. The challenge lasted 122 days and attracted 1,616 submissions from 112 participants on the tuning set evaluation platform, resulting in a total of 58,100 predictions. After the main challenge event at MICCAI, we collected three new cohorts and conducted the post-challenge analysis, aiming to evaluate the generalization ability of the top-performing algorithms.

\begin{table}[!t]
\caption{\textbf{Dataset characteristics of the development set, internal validation set, and three external validation sets.} The development set includes a training set with 50 labeled images and 2000 unlabeled images and a tuning set with 50 images. All the images in the development set are available to participants but the tuning set annotations are not released. All validation sets are fully independent and hidden from participants. For each organ, we report the statistics of volume or diameter. Values are displayed as Median values (First quartile, Third quartile).}
\centering
\resizebox{\textwidth}{!}{
\begin{tabular}{llllll} 
\hline
Name                & \multicolumn{1}{c}{Development set}                                                                     & \multicolumn{1}{c}{\begin{tabular}[c]{@{}c@{}}Internal \\ validation set\end{tabular}}                                                                                 & \multicolumn{1}{c}{\begin{tabular}[c]{@{}c@{}}North American\\ external validation set\end{tabular}} & \multicolumn{1}{c}{\begin{tabular}[c]{@{}c@{}}European external \\ validation set\end{tabular}} & \multicolumn{1}{c}{\begin{tabular}[c]{@{}c@{}}Asian external\\ validation set\end{tabular}}      \\ 
\hline
No. of cases        & \multicolumn{1}{c}{2100}                                                                                & \multicolumn{1}{c}{200}                                                                                                                                                & \multicolumn{1}{c}{200}                                                                         & \multicolumn{1}{c}{200}                                                                         & \multicolumn{1}{c}{200}                                                                          \\ 
\hline
No. of sources      & \multicolumn{1}{c}{22}                                                                                  & \multicolumn{1}{c}{23}                                                                                                                                                 & \multicolumn{1}{c}{2}                                                                           & \multicolumn{1}{c}{2}                                                                           & \multicolumn{1}{c}{4}                                                                            \\ 
\hline
Region                & \begin{tabular}[c]{@{}l@{}}North American\\ European\end{tabular}                                       & \begin{tabular}[c]{@{}l@{}}North American\\ European\end{tabular}                                                                                                      & North American                                                                                  & European                                                                                        & Asian                                                                                            \\ 
\hline
Disease             & \begin{tabular}[c]{@{}l@{}}Liver cancer\\ Kidney cancer\\ Spleen disease\\ Pancreas cancer\end{tabular} & \begin{tabular}[c]{@{}l@{}}Liver cancer\\ Kidney cancer\\ Spleen disease\\ Pancreas cancer\\ Stomach cancer\\ Sarcomas\\ Ovarian cancer\\ Bladder disease\end{tabular} & \begin{tabular}[c]{@{}l@{}}Liver cancer\\ Pancreas cancer\end{tabular}                          & \begin{tabular}[c]{@{}l@{}}Various \\ abdominal \\ disease\end{tabular}                         & \begin{tabular}[c]{@{}l@{}}Various \\ abdominal \\ disease\end{tabular}                          \\ 
\hline
Phase               & \begin{tabular}[c]{@{}l@{}}Plain phase\\ Artery phase\\ Portal phase\\ Delay phase\end{tabular}         & \begin{tabular}[c]{@{}l@{}}Plain phase\\ Artery phase\\ Portal phase\\ Delay phase\end{tabular}                                                                        & \begin{tabular}[c]{@{}l@{}}Plain phase\\ Artery phase\\ Portal phase\\ Delay phase\end{tabular} & \begin{tabular}[c]{@{}l@{}}Plain phase\\ Artery phase\\ Portal phase\\ Delay phase\end{tabular} & \begin{tabular}[c]{@{}l@{}}Plain phase\\ Artery phase\\ Portal phase\\ Delay phase\end{tabular}  \\ 
\hline
Manufacturer              & \begin{tabular}[c]{@{}l@{}}Siemens, GE, \\Philips,~Toshiba, \\ Barco, Vital\end{tabular}                & \begin{tabular}[c]{@{}l@{}}Siemens, GE, \\Philips, Toshiba,\\Imatron, Vital\\PHMS\end{tabular}                                                                         & \begin{tabular}[c]{@{}l@{}}GE,Philips, \\Toshiba\end{tabular}                                   & \begin{tabular}[c]{@{}l@{}}Siemens,~GE,\\Philips, Toshiba,\end{tabular}                         & \begin{tabular}[c]{@{}l@{}}Siemens,~GE,\\Philips, Toshiba,\end{tabular}                          \\ 
\hline
Liver volume        & 1601 (1383, 1921)                                                                                       & 1542 (1323, 1655)                                                                                                                                                      & 1611 (1264, 1970)                                                                               & 1458 (1223, 1662)                                                                               & 1188 (1013, 1367)                                                                                 \\
Right kidney volume & 197.6 (162.4, 254.5)                                                                                    & 179.0 (152.5, 207.5)                                                                                                                                                   & 170.6 (149.4, 212.7)                                                                            & 158.7 (125.2, 196.1)                                                                            & 141.8 (122.6, 161.9)                                                                             \\
Left kidney volume  & 196.6 (166.8, 251.1)                                                                                    & 184.4 (146.8, 213.1)                                                                                                                                                   & 175.8 (153.6, 201.0)                                                                            & 153.7 (126.3, 193.1)                                                                            & 146.7 (126.5, 174.9)                                                                             \\
Spleen volume       & 230.7 (158.8, 320.3)                                                                                    & 230.3 (174.9, 308.4)                                                                                                                                                   & 292.2 (164.4, 497.3)                                                                            & 182.3 (133, 271.4)                                                                            & 182.9 (130.0, 252.4)                                                                             \\
Pancreas volume     & 89.16 (67.96, 119.3)                                                                                    & 77.38 (64.34, 92.93)                                                                                                                                                   & 76.14 (59.17, 90.09)                                                                            & 73.19 (58.55, 92.61)                                                                           & 74.26 (60.43, 91.70)                                                                             \\
Aorta diameter      & 21.83 (20.40, 24.45)                                                                                    & 23.24 (21.53, 25.44)                                                                                                                                                   & 23.66 (21.30, 25.85)                                                                            & 24.32 (21.41, 27.08)                                                                            & 21.62 (19.42, 24.04)                                                                             \\
Inferior vena cava  & 26.12 (24.43, 27.90)                                                                                    & 26.01 (23.56, 27.84)                                                                                                                                                   & 25.19 (23.63, 27.41)                                                                            & 25.83 (23.67, 27.71)                                                                            & 24.79 (23.03, 26.39)                                                                             \\
Right adrenal gland & 4.222 (3.551, 5.540)                                                                                    & 4.802 (3.794, 5.573)                                                                                                                                                   & 4.340 (3.350, 5.339)                                                                            & 3.829 (2.747, 4.994)                                                                            & 3.297 (2.513, 4.342)                                                                             \\
Left adrenal gland  & 5.265 (4.348, 6.404)                                                                                    & 5.423 (4.342, 6.474)                                                                                                                                                   & 4.963 (3.938, 6.137)                                                                            & 4.195 (3.012, 5.564)                                                                           & 3.977 (3.187, 5.048)                                                                             \\
Gallbladder volume  & 26.80 (16.92, 44.33)                                                                                    & 28.94 (17.66, 41.78)                                                                                                                                                   & 23.53 (14.04, 44.12)                                                                            & 22.5 (12.08, 34.27)                                                                            & 18.17 (8.734, 27.76)                                                                             \\
Esophagus diameter  & 19.11 (17.16, 21.66)                                                                                    & 18.55 (16.70, 19.87)                                                                                                                                                   & 20.49 (18.09, 22.72)                                                                            & 18.93 (16.54, 21.13)                                                                            & 17.85 (15.83, 20.57)                                                                             \\
Stomach volume      & 308.7 (226.9, 419.6)                                                                                    & 349.1 (256.3, 525.4)                                                                                                                                                   & 371.2 (274.1, 510.5)                                                                            & 310.7 (208.2, 472.9)                                                                            & 342.3 (214.8, 502.2)                                                                             \\
Duodenum diameter   & 76.52 (65.93, 87.74)                                                                                    & 71.20 (61.71, 89.64)                                                                                                                                                   & 74.32 (64.90, 88.99)                                                                            & 65.61 (52.55, 81.82)                                                                            & 58.23 (48.28, 72.93)                                                                             \\
\hline
\end{tabular}}
\end{table}

\subsection*{Dataset characteristic}
To create a comprehensive representation of real-world scenarios, we curated a robust and diverse dataset sourced from 53 distinct medical groups. This dataset encompassed a wide range of attributes, including varying racial backgrounds, multiple diseases, different imaging phases, and diverse CT machine manufacturers (Fig.~\ref{fig:overview}c, Table 1, Supplementary Table 1-3). In total, 2900 images were retrospectively de-identified for the algorithm development and independent validation,
% . Specifically, the development set contains 2100 cases (2000 unlabeled cases and 50 labeled cases in the training set, 50 labeled cases in the tuning set) from 22 medical groups, 
covering North American and European patients with typical abdominal diseases: liver cancer, kidney cancer, spleen disease, and pancreas cancer. 
The internal validation set consists of 200 cases where 100 cases with the same disease type as the development set and 100 cases with new abdominal diseases, such as stomach cancer, sarcomas, colon cancer, ovarian cancer, and bladder diseases. 
The external validation sets are from three individual cohorts with North American, European, and Asian patients, respectively. Each external validation set contains 200 labeled cases from multiple medical centers. Annotation quality is highly important for model performance~\cite{Lena-Annotation} and we employed a hierarchical human-in-the-loop pipeline to maintain consistent annotations (Methods).

\subsection*{Overview of evaluated algorithms}
We received 48 successful algorithm docker containers during the validation phase. Each algorithm docker container was independently evaluated on the same workstation to obtain five quantitative metrics on the internal validation set and the final rank was generated with the `rank-then-aggregate' strategy (Methods). 
The evaluation metrics and the ranking scheme has been made available to participants at the beginning of the challenge for the sake of fairness and transparency~\cite{Lena2018rankings-NC}. The top 10 teams were invited to join the FLARE consortium for post-challenge analysis and ablation study using unlabeled data. Four teams ranked top 10 in terms of segmentation accuracy but were not ranked top 10 overall because of low efficiency. However, in some clinical scenarios where algorithm efficiency is not the major consideration, such algorithms are also useful. Therefore, we also invited these teams to join the FLARE consortium and conducted the post-challenge analysis. We refer to the algorithms of the 14 teams as the top-performing algorithms. 

We analyzed the characteristics of the employed algorithms by participants (Fig.~\ref{fig:overview}e-g, Table 4-5).
All the submitted algorithms were based on deep learning. 80\% teams used U-Net as the main network architecture (Fig.~\ref{fig:overview}e), where 45\% teams used plain U-Net and others combined U-Net with popular networks, such as ResNet (23\%) and Transformer~\cite{ViT2020}. 
All the algorithms used dice loss~\cite{milletari2016vnet} and its combination with cross-entropy loss was the most popular choice (Fig.~\ref{fig:overview}f) because compound loss function has been proven to be robust~\cite{SegLossOdyssey}.  
Adam~\cite{kingma2014adam} was the most frequently used optimizer followed by the stochastic gradient descent (SGD) (Fig.~\ref{fig:overview}g).

\textbf{Best-performing algorithm.} Team blackbean (T1~\cite{blackbean-Flare22-Design}) proposed a pseudo-labelling framework with nnU-Net~\cite{isensee2020nnunet} as the base network architecture. Multiple big nnU-Net models were trained based on the labeled cases. Then, unlabeled cases were passed to the trained models to predict pseudo-labels. After that, the big nnU-Net models were updated based on the combination of labeled cases and unlabeled cases with pseudo-labels.  The process was iterated for three rounds and low-quality pseudo-labels were filtered by the uncertainty score which was defined based on the changes in pseudo-labels during different iterations. To improve the inference efficiency, a small nnU-Net was trained by reducing the input size and network size. Furthermore, a modified sliding window strategy was designed to filter background regions based on anatomy prior, which can reduce the computational cost.

\begin{figure*}[!htbp]%
\centering
\includegraphics[scale=0.335]{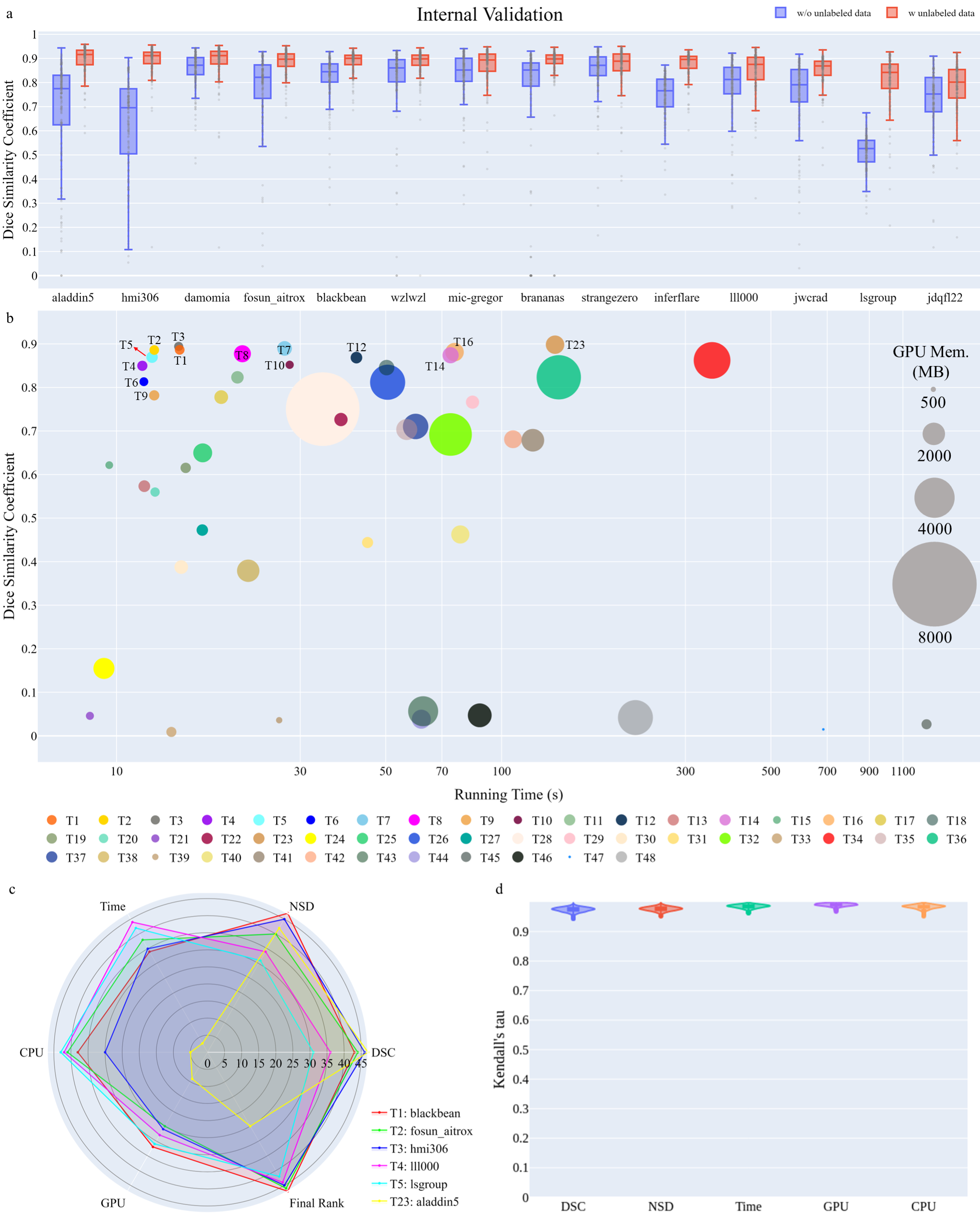}
\caption{\textbf{Performance analysis on the internal validation set.} \textbf{a,} The comparisons of using unlabeled data and without using unlabeled data for top-performing algorithms show that using unlabeled data can significantly improve the performance. The average improvement of the Dice Similarity Coefficient (DSC) score is 9.8\%. \textbf{b,} The top three best-performing algorithms achieve a good trade-off between segmentation accuracy (y-axis) and efficiency (x-axis). The circle size is proportional to GPU memory consumption. The 14 top-performing algorithms are marked in the figure. \textbf{c,} The performance comparisons among different dimensions are presented for the top three best-performing algorithms and another three top-performing algorithms with the best DSC, running time, and CPU utilization metrics, respectively. The value denotes the number of teams surpassed by each algorithm in each dimension. \textbf{d,} The bootstrap distribution of rankings (N=1000) shows that the ranking scheme is stable with respect to sampling variability.
}\label{fig:interval}
\end{figure*}

\textbf{Best-segmentation-accuracy algorithm.} Team aladdin5 (T23~\cite{aladdin5-Flare22-Design}) developed a cascaded framework with nnU-Net~\cite{isensee2020nnunet} and pseudo-label learning. A binary U-Net segmentation network was first trained to localize the ROI in low-resolution images. Then, an augmented U-Net with more trainable parameters was used to segment the 13 target organs from the ROI in high-resolution images. In addition, an ensemble of multiple U-Nets was used to segment the unlabeled images to generate pseudo-labels. Finally, the cascaded framework was trained with 50 labeled cases and 2000 cases with pseudo-labels. However, this framework is time-consuming because of the large model size. The inference speed was nearly 10 times slower than the top three best-performing teams. 

% All top-performing teams in the challenge shared two prominent features in their algorithms. First, a U-Net-like encoder-decoder architecture was employed as the foundation of their network architectures. This architecture can preserve large contextual information and extract multi-level and multi-scale features from input images, which played a vital role in enhancing segmentation performance. Second, all these leading teams embraced the utilization of pseudo-label learning to explore valuable knowledge from unlabeled images. This strategic choice demonstrated the efficacy of utilizing unlabeled images to augment segmentation performance, which holds great potential for extension to other medical image segmentation tasks where annotated data is scarce, given the abundance of unlabeled images typically available in various medical scenarios.

% A summary of the other top-performing algorithms is presented in Supplementary Table TBD.
% We invited all the top-performing algorithms to make their code and docker containers publicly available to the research community (Supplementary Table 4). These valuable resources offer researchers the opportunity to train customized models on new datasets, providing a foundation for further advancements in the field. Additionally, the availability of docker containers streamlines the process of generating abdominal organ segmentations for new CT scans, eliminating the need for extensive configuration of deep learning environments. 

\subsection*{Performance analysis on the internal validation dataset}
We independently evaluated all the submitted algorithms on the holdout internal validation set during the validation phase. Dice Similarity Coefficient (DSC) and Normalized Surface Distance (NSD)~\cite{maier2022metrics} were used for quantitative evaluation. As shown in Fig.~\ref{fig:interval}a and Supplementary Table 6, the algorithms with the best DSC scores demonstrated remarkable agreement with the organ contours of the ground truth. The median average organ DSC score is 91.6\% (interquartile range (IQR): 87.4-93.4\%), significantly surpassing the performance of other teams (Wilcoxon signed-rank test, $p  < 0.05$).

To evaluate the impact of unlabeled data, we conducted a comparison between the algorithms utilizing unlabeled data and their counterparts that did not employ the unlabeled data (Fig.~\ref{fig:interval}a). Notably, all algorithms exhibited significant performance improvements (paired T-test $p < 0.05$) when leveraging unlabeled images. The team hmi306 achieved the highest performance gain of 26.8\%.
% while the average performance gain across all teams was 9.8\% in terms of DSC. 
Furthermore, the incorporation of unlabeled cases contributed to a reduction in the number of cases with poor segmentation. These compelling results demonstrate that the use of unlabeled data has the potential to enhance model performance and improve generalization capabilities.
% , particularly when there is a scarcity of annotated data.

Algorithm efficiency is another important metric to consider when deploying models. We quantitatively evaluated the running time and resource usage of all algorithms (Fig.~\ref{fig:interval}b, Methods, Supplementary Table 6, Fig. 2). For an input image with more than twenty million voxels ($\sim 512\times512\times100$), all the top ten algorithms (T1-T10) can finish the segmentation of 13 organs within 30 seconds. As depicted in the bubble plot (Fig.~\ref{fig:interval}b), the majority of top-performing teams aimed to strike a balance between computing resource consumption and accuracy, evidenced by their concentration towards the upper left of the figure. 
% However, some teams focused on optimizing a single aspect of the algorithms, which resulted in a lower overall ranking. For example, while team T21 favored rapid inference speed with limited computing resources, their segmentation accuracy was subpar. Conversely, team T16 produced commendable DSC scores, albeit at the expense of considerably longer inference times and greater GPU memory usage.
Remarkably, the top three teams exhibited an excellent trade-off between segmentation accuracy and efficiency, achieving a median DSC of 88.6-89.4\% in under 15 seconds, with average GPU memory usage below 2GB and CPU utilization below 30\%. 
% These teams also demonstrated reduced resource consumption, 
These findings highlight the notable efficiency achieved by the leading algorithms, making them well-suited for practical model deployments.

% We further analyzed the characteristics exhibited by the top-performing algorithms, specifically T1: blackbean, T2: fosun\_aitrox, T3: hmi306, as well as three algorithms with outstanding performance in individual metrics: T23: aladdin5 (best DSC), T4: lll000 (fastest running time), and T5: lsgroup (lowest CPU utilization) (Fig.~\ref{fig:interval}c). To highlight these characteristics, we employed a radar plot that captured six dimensions for each algorithm. 
We further analyzed the characteristics exhibited by the top-performing algorithms with a radar plot that captured six dimensions for each algorithm (Fig.~\ref{fig:interval}c). It revealed that algorithms solely prioritizing a single metric, such as DSC, failed to achieve top rankings. In contrast, algorithms striking a favorable trade-off between segmentation accuracy and efficiency secured prominent positions in the rankings. 
We also evaluated the ranking stability by performing bootstrap (1000 times) for all algorithms and computed Kendall's $\tau$ as a quantitative metric (Fig.~\ref{fig:interval}d, Methods). For each metric, Kendall's $\tau$ scores were found to be very close to 1, indicating a high degree of agreement between the rankings. Additionally, the compact distributions of these scores further confirm the stability of the ranking results with respect to sampling variability. These findings provide robust evidence that the obtained rankings are highly consistent and reliable across different samples.

% : DSC, NSD, running time, CPU utilization, GPU consumption, and the final rank on the internal validation set. In each dimension, we determined the number of teams surpassed by each algorithm. 
% Our analysis revealed that algorithms solely focused on segmentation accuracy did not achieve top rankings. For example, while team aladdin5 attained the best DSC score, its segmentation efficiency metrics fell behind other teams, resulting in a relatively average overall rank. In contrast, teams fosun\_aitrox and hmi306 demonstrated an exceptional trade-off between segmentation accuracy and efficiency, leading to their prominent positions (2nd and 3rd place, respectively) in the final rankings, despite not achieving the best performance in all metrics.

% In order to evaluate the ranking stability, we performed bootstrap with 1000 times for all algorithms and computed Kendall's $\tau$ as a quantitative metric (Fig.~\ref{fig:interval}d, Methods). 
% The ranking list based on the full assessment interval validation set was pairwise compared with the ranking lists based on the individual bootstrap samples. The median Kendall's $\tau$ scores were 0.990 (IQR: 0.986-0.994) for GPU consumption, 0.986 (IQR: 0.982-0.989) for running time, 0.984 (IQR: 0.979-0.989) for CPU utilization, 0.977 (IQR: 0.972-0.982) for NSD score, and 0.975 (IQR: 0.968-0.979) for DSC score. 
% To illustrate these results, violin plots were used to depict the distribution of Kendall's $\tau$ scores 

\begin{figure*}[!htbp]%
\centering
\includegraphics[scale=0.33]{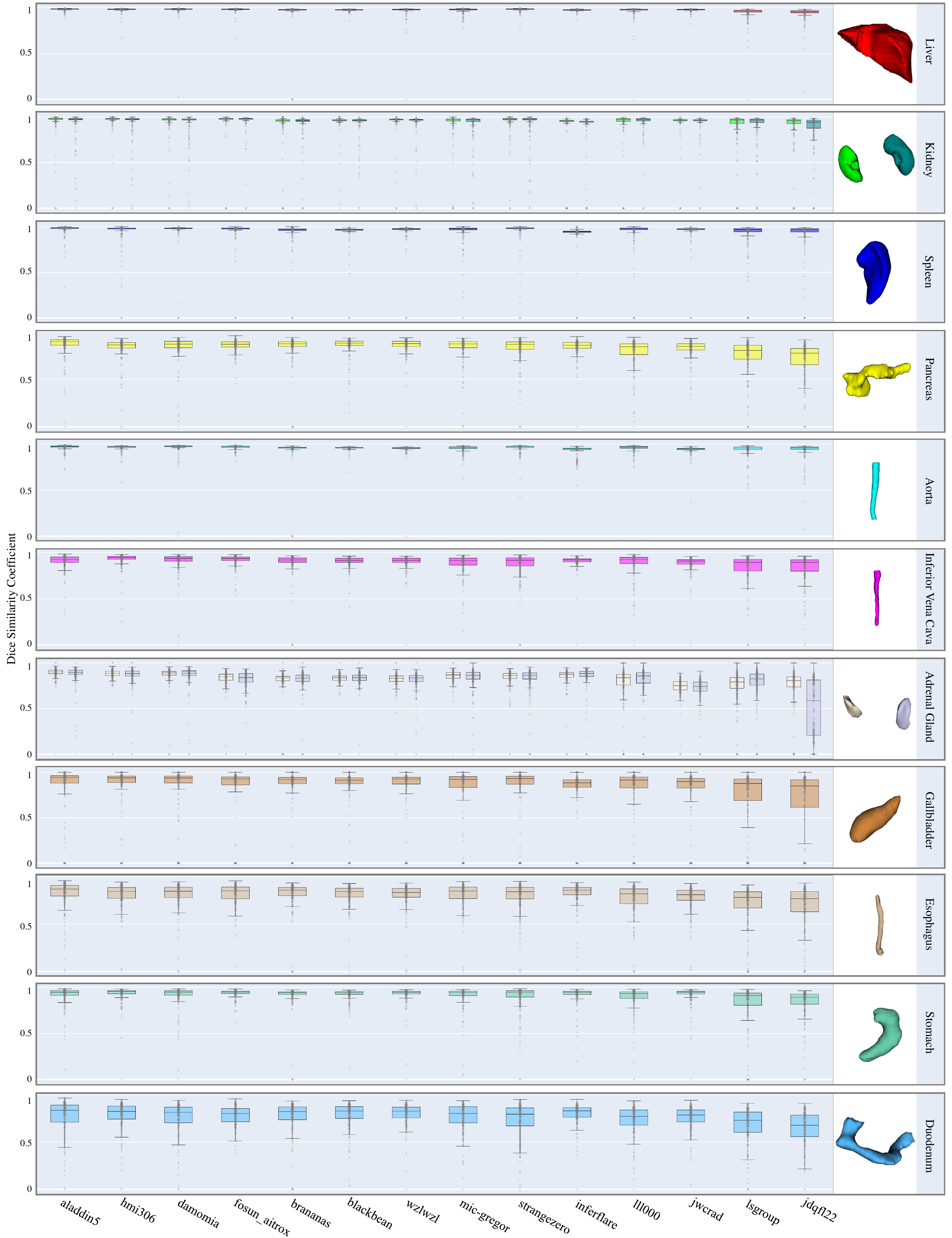}
\caption{\textbf{Dot and box plots of the Dice Similarity Coefficient (DSC) values of top-performing algorithms for the 13 organs on the interval validation set.} The box plots display descriptive statistics across all internal validation cases, with the median value represented by the black horizontal line within the box, the lower and upper quartiles delineating the borders of the box, and the vertical black lines indicating the 1.5 interquartile range. The algorithms are ranked on the x-axis based on their median DSC scores. 
}\label{fig:DSC-Organ-Interval}
\end{figure*}

Finally, we conducted an in-depth analysis of the organ-wise segmentation performance exhibited by the 14 top-performing algorithms (Fig.~\ref{fig:DSC-Organ-Interval}). 
Notably, all the top algorithms achieved high DSC scores of over 90\% for organs such as the liver, kidneys, and spleen, owing to their relatively larger volumes and simple shapes, which make them easier to segment accurately. For example, the best-segmentation-accuracy team aladdin5 achieved median DSC scores of 98.6\% (IQR: 98.3-98.8\%), 98.3\% (IQR: 97.3-98.7\%), 98.2\% (IQR: 97.0-98.5\%), and 98.6\% (IQR: 98.0-99.0\%) for the liver, right kidney, left kidney, and spleen, respectively. 

In the case of tubular organs, the aorta and inferior vena cava (IVC) showed higher medical DSC scores than the esophagus. This discrepancy can be attributed to the lower contrast and thinner caliber of the esophagus, making it more challenging to segment.
% Team hmi306, damomia, and aladdin5 achieved the best segmentation accuracy for the aorta, IVC, and esophagus, respectively, with median DSC scores of 97.7\% (IQR: 97.2-98.0\%), 94.7\% (IQR: 92.7-96.0\%), and 86.9\% (IQR: 79.7-91.0\%).
The segmentation of the remaining organs poses the greatest challenge. Specifically, the pancreas, stomach, and duodenum exhibit complex structures with boundaries that may overlap with surrounding organs. The adrenal glands and gallbladder, being relatively small within the abdomen, are also difficult to accurately recognize. These organs may be surgically removed in some cases, further complicating the segmentation task. As a result, the algorithms showcased varied performance when it came to these challenging organs. Notably, team aladdin5 achieved the best accuracy for these difficult organs, with median DSC scores ranging from 83.7\% to 94.4\%. Altogether, the top-performing algorithms demonstrated high segmentation accuracy for organs such as the liver, kidneys, and spleen, while facing challenges with more complex structures, smaller organs, and surgically altered organs, where diverse performance was observed.

\begin{figure*}[!htbp]%
\centering
\includegraphics[scale=0.4]{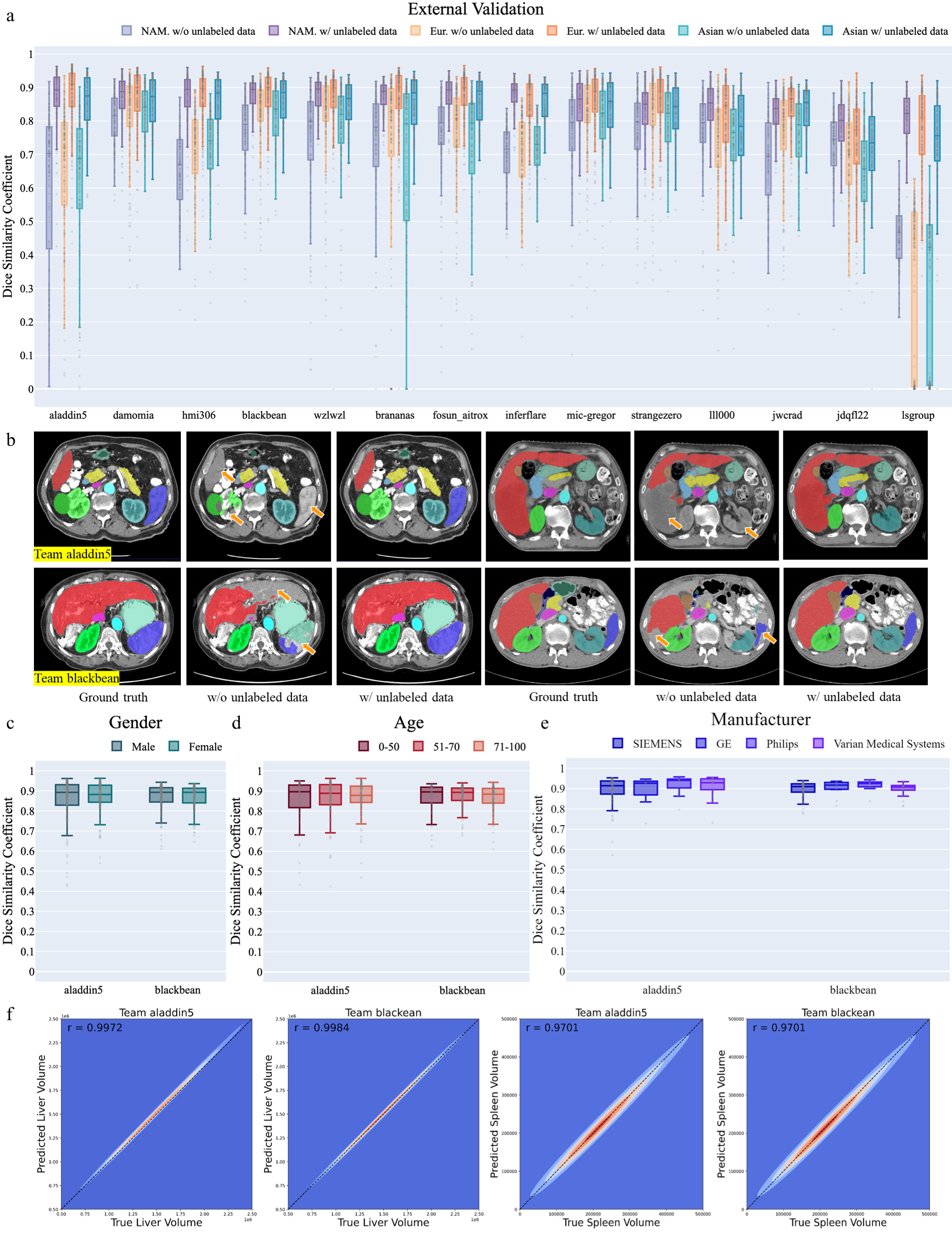}
\caption{\textbf{Performance on three external validation sets.} \textbf{a,} The segmentation performance (Dice similarity coefficient, DSC) on the North American (NAM.) cohort, European (Eur.) cohort, and Asian cohort. For each cohort, the DSC scores between using unlabeled data and without using unlabeled data are presented as well. \textbf{b,} Visualized segmentation examples of the two top algorithms show that using unlabeled data can significantly improve the segmentation quality. \textbf{c-e,} The segmentation performance of the best-accuracy algorithm (aladdin5) and the best-performing algorithm (blackbean) across demographics, including genders, ages, and manufacturers. \textbf{f,} Pearson's correlation contour plots of the organ volume demonstrate that the two top algorithms accurately quantify the liver and spleen volumes, which are important clinical biomarkers. 
}\label{fig:externalVal}
\end{figure*}

\subsection*{Performance analysis on the external validation cohorts}
One of the primary challenges limiting the widespread application of AI in clinical routines is the lack of generalization capability to new cohort data. To quantitatively assess the generalization performance, we conducted an independent evaluation of the top 14 algorithms on three external validation cohorts: the North American, European, and Asian cohorts (Fig.~\ref{fig:externalVal}), which are curated from external medical centers (Supplementary Table 3). 
% Additionally, we evaluated the performance of the corresponding algorithms without utilizing unlabeled data, demonstrating the performance gains through unlabeled data.
% These cohorts were curated from new medical centers that had no overlap with the previous development and internal validation sets (Supplementary Table 3). Additionally, we evaluated the performance of the corresponding algorithms without utilizing unlabeled data, enabling a quantitative analysis of the performance gains achieved through the use of unlabeled data.

The best-accuracy algorithm (aladdin5) generalized well on the external validation cohorts (Fig.~\ref{fig:externalVal}a, Supplementary Fig. 3b, Table 6-8). In comparison to the internal validation set, the algorithm achieved comparable performance with median DSC scores of 89.3\% (IQR: 84.4-93.0\%), 90.9\% (IQR: 84.3-94.2\%), and 87.5 (IQR: 80.3-92.9\%) on the North American, European, and Asian cohorts, respectively. 
%The performance of the Asian cohort was slightly lower than other cohorts. This probably due to the data distribution shift between the training set and the external Asian cohort while the development set mainly contained North American and European images. Nevertheless, the performance gap between different cohorts is marginal, indicating that the top algorithm can generalize to unseen images from new cohorts.
Among the 14 top algorithms, multiple algorithms obtained competitive performance compared to the best DSC score,
%On the North American cohort, seven algorithms achieved a median DSC score of 88.8-89.5\%, six algorithms achieved a median DSC score of 89.8-90.9\% on the European cohort, and five algorithms reached a median DSC score of 87.3-89.0\% on the Asian cohort. 
%All three best-performing algorithms (T1: blackbean, T2: fosun\_aitrox, and T3: hmi306) still achieved high performance on the three external cohorts, 
indicating their strong generalization ability on the external validation sets.
Furthermore, organ-wise analysis (Supplementary Fig. 4-9) shows that these algorithms can accurately segment most organs (e.g., liver, kidneys, spleen, and aorta) with median DSC and NSD scores of more than 90\%. However, for some challenging cases, such as organs with low contrast or severe pathological changes, there is still large room for further improvements (Fig. 10-12).  
Moreover, compared to the counterpart algorithms without unlabeled data, the top algorithms using unlabeled data still consistently achieved remarkable performance gains on the three external cohorts, indicating that using the unlabeled data during the development phase can significantly improve the model generalization ability.
% Overall, the average DSC score improvements are 10.2\%, 9.2\%, and 11.3\% on the North American, European, and Asian cohorts, respectively, indicating that using the unlabeled data during the development phase can significantly improve the model generalization ability. This demonstrates that unlabeled data could be used to reduce the annotation requirements for AI algorithms.  

In order to understand how the unlabeled data improves the segmentation quality, we visualized typical segmentation examples from the best-accuracy algorithm (T23: aladdin5) and the best-performing algorithm (T1: blackbean)  (Fig.~\ref{fig:externalVal}b, Supplementary Fig. 13-15). It can be found that the baseline model (without using unlabeled data) is not robust to different image contrasts, generating various over-segmentation and under-segmentation errors. 
% For example, the liver and spleen were missed and the right kidney with a cyst was under-segmented in the top-left example. 
% In the remaining examples, the liver segmentation masks were not complete and the spleen was miss-segmented as the stomach in the bottom-left example because of the low image contrast and touched boundary. 
In contrast, these segmentation errors were significantly reduced by learning with unlabeled data, demonstrating that unlabeled data can be used to improve the model generalization ability. 

AI models have been proven that they could have biases on different genders, ages, and other demographic factors~\cite{FairAI-NMed-18}. Therefore, it is necessary to evaluate the fairness of the top algorithms. 
We evaluated the fine-grained performance of the best-accuracy algorithm (T23: aladdin5) and the best-performing algorithm (T1: blackbean) across different genders, ages, and manufacturers on the external validation set. 
We separate gender groups to male and female, age groups to 0-50, 51-70, and over 70, manufacturer groups to SIEMENS, General Electric (GE), Philips, and Varian Medical System. The results imply that the two algorithms do not significantly vary regarding gender (Fig.~\ref{fig:externalVal}c and Supplementary Fig. 3c), age (Fig.~\ref{fig:externalVal}d and Supplementary Fig. 3d), and manufacturer (Fig.~\ref{fig:externalVal}e and Supplementary Fig. 3e), highlighting their potential for broad applications in abdominal CT segmentation tasks. 
Moreover, the resulting algorithms enable automatic and high-throughput image-based phenotyping. We computed liver and spleen volume based on the segmentation results of algorithms aladdin5 and blackbean (Fig. ~\ref{fig:externalVal}f). There is a strong correlation (Pearson's r=0.9701-0.9984) between the predicted and ground-truth volume. This indicates that the two top algorithms accurately quantify organ volumes, which could replace manual linear measurements.

\section*{Discussion}\label{diss}
% P1. Unsolved problems and study objectives
% AI has revolutionized medical image analysis tasks (e.g., quantification, diagnosis, and grading), but most algorithms rely on a large number of human expert annotations which are extremely hard and expensive to collect. Moreover, the performance of existing algorithms is mainly evaluated based on accuracy-related metrics on limited cohorts while the generalization ability, running efficiency, and resource consumption are overlooked. These barriers hinder the wider adoption of AI algorithms in clinical practice. The main goal of this study was to address these critical issues. First, we created a large and divese abdomen CT dataset with a well-defined segmentation task to benchmark algorithms. Second, we organized an international competition to gather community efforts to facilitate AI algorithm developments for learning with unlabeled data. Third, we validated the algorithms in real-world settings by blinded evaluation of their generalization ability, efficiency, and resource consumption on intercontinental and multinational cohorts. 

% P2. Challenge achievements
The FLARE 2022 challenge presented the largest challenge in abdomen image analysis in terms of both dataset size and the number of docker-packed algorithm submissions. The datasets covered real-world diversities of races, patients, abdominal cancer types, imaging manufacturers, and imaging protocols. The challenge attracted participants from all over the world with different research backgrounds, such as computer vision, biomedical engineering, biology, and radiology. Moreover, this was the first documented attempt to analyze the usage of unlabeled data for boosting segmentation performance and quantitative evaluation of algorithm efficiency and resource consumption. 
The resulting algorithms may reduce the time to extract abdominal organ biomarkers, empowering radiologists and clinicians to adopt and conduct more quantitative analysis for research and clinical practice.

% P3. Main findings
There are three main findings based on the validation results. First, unlabeled data can be used to alleviate the annotation shortage problem and significantly improve algorithm generalization ability.  
Most of the participants employed similar network architectures: U-Net-like fully convolutional networks. The main factor to distinguish the best-performing teams was the successful exploitation of unlabeled data with pseudo-label learning. 
Second, the results demonstrate that AI algorithms can achieve a great trade-off between accuracy, efficiency, and resource consumption. 
% For example, team aladdin5 achieved the best segmentation accuracy with an average DSC and NSD of 89.8\% and 93.7\% on the internal validation set, while team blackbean's approach obtained comparable accuracy (DSC 88.6\% and NSD 96.7\%) but was 9$\times$ faster and reduced GPU memory consumption and CPU utilization by 93\% and 88\%, respectively.
Third, most organs have obtained very high agreements with reference standards but some organs still have a large room for further improvements, such as the pancreas, stomach, and duodenum because of their irregular shapes and dynamic appearances. This indicates that abdominal organ segmentation is still an unsolved problem (Supplementary Fig. 10-12).

% P4.1 Internal validation set performance (common strategies for winners)
We also identify some useful strategies on how to use unlabeled data to improve algorithm performance. All the best-performing teams used pseudo-label learning to incorporate the unlabeled data into the algorithm development.  The main idea is to train models on a small well-labeled dataset and predict the unlabeled data to generate pseudo-labels. Then, one can update the model weights with all the data. In this pipeline, the key is to generate high-quality pseudo-labels. Participants have developed different strategies to improve the quality of pseudo-labels. For example, team aladdin5 generated pseudo-labels by model ensembles which can incorporate knowledge from different models, while team blackbean defined an uncertainty metric to select confident pseudo-labels. By contrast, team fosun\_aitrox employed an easy-to-hard curriculum learning manner to prioritize reliable pseudo-labels. The resulting algorithms have been extensively validated on the internal and external validation sets and demonstrated their effectiveness to bring remarkable performance gains. In addition, data augmentation was also used by all top teams, such as rotation, scaling, and intensity shifting. This augmentation transforms can improve algorithms' robustness to different CT imaging protocols. 

% P4.2 Strategies for fast inference speed: two-stage framework; whole volume input; image priors
Resource consumption and efficiency are also important factors during algorithm deployments in clinical environments because most medical centers do not have powerful computing devices. 
% Thus, we quantitatively evaluated the running time, GPU memory consumption, and CPU utilization for submitted algorithms and included these metrics in the ranking scheme. Furthermore, we identify commonly used strategies to reduce resource consumption and improve inference speed. Specifically, 
Many top teams employed two-stage frameworks. In the first stage, a small network was used roughly locate the abdomen region of interest (ROI). In the second stage, a larger network was used for fine-grained segmentation. To further improve the inference speed, one can resize the whole CT scan to a fixed image size in the first stage, which can reduce the number of voxels by 20 times. 
% This operation could lose image details but the remaining information is still enough to locate the abdomen ROI. 
Moreover, image priors can also be used to reduce the computational burden. For example, the human tissues are located in the middle of the CT scans and the background intensities are zero. Thus, the algorithm prediction process can directly ignore the background and only predict the middle part of the input CT image.  

% P6 challenge design comments
% We aimed to stimulate the exploration of unlabeled medical image data by providing a carefully curated dataset and organizing a global challenge in which participants can compare their algorithms in a fair platform. The resulting algorithms can simultaneously segment multiple organs with a great trade-off between accuracy and efficiency and generalize well on multiple external validation cohorts, indicating that the well-designed challenge can lead to rapid algorithm development. The top three algorithms can finish the inference process within 5 seconds for typical 3D CT scans with millions of voxels using less than 2GB GPU memory ($\textless$ 100 USD), which is affordable in clinical practice. We also avoid some typical pitfalls in medical AI challenges~\cite{Lena2018rankings-NC}, such as the limited reproducibility and interpretation of the challenge results and the low stability of the ranking results. 
% These pitfalls were addressed during the designing stage of the FLARE challenge. We registered the challenge following the official BIAS guidelines and made the proposal document publicly available. The evaluation metrics, code, and ranking scheme are also available at the beginning of the challenge. In addition, we conducted a comprehensive post-challenge analysis to validate where the top algorithms can generalize to unseen new cohorts. 

% P7 limitation and future work
This study has two main limitations. First, although we have constructed the largest abdomen CT dataset, the included patients were predominantly North American, European, and Asian, which lack patients from Africa. Second, this study mainly focuses on organ quantification while tumor quantification is also important in clinical practice. In the near future, we will increase the tumor annotations for this dataset and extend the challenge with joint organ and tumor segmentation. 
% Moreover, data availability is a limitation for achieving good performance and generalization ability for most medical image analysis tasks. We hope others in the wider medical imaging community will be interested in contributing more images to further enlarge this resource, including cases for new medical centers and races, especially for the underrepresented cohorts. Although annotated images are desired, this study has demonstrated that even unlabeled images are useful.

% P8 conclusion
This paper presents a successful proof of concept that unlabeled data are indeed useful in AI algorithm development when annotations are in shortage. The top algorithms obtained consistent and significant improvement by using the unlabeled data. Moreover, abdominal organ segmentation is still an unsolved problem, especially for organs with complex shapes and diverse appearances. We anticipate that in the near future, radiologists and clinicians can be assisted by AI algorithms to objectively and automatically assess organ status from abdominal CT scans at high throughput. In addition, we have made this large abdominal dataset and the code repositories of top algorithms publicly available to the community for research use \url{https://flare22.grand-challenge.org/}. This open-source commitment not only facilitates reproducibility but also encourages collaboration and fosters progress in abdominal disease research.

% P5 External performance
% failure mode,
% reason: data distribution shift
% potential solution: calibrate model with target center data
% algorithms seg features
% unsolved problem: the semi-supervised algorithms mainly learn general information from the unlabeled images while some rare diseases or low image quality are still unsolved problems.

% Conclusions may be used to restate your hypothesis or research question, restate your major findings, explain the relevance and the added value of your work, highlight any limitations of your study, describe future directions for research and recommendations. 
% In some disciplines use of Discussion or 'Conclusion' is interchangeable. It is not mandatory to use both. Please refer to Journal-level guidance for any specific requirements. 

\subsection*{FLARE challenge consortium} 
Junjun He, Hua Yang, Huihua Yang, Bingding Huang, Mengye Lyu, Yongkang Ma, Heng Guo, Rongguo Zhang, and Klaus Maier-Hein \\
Jun Ma is with the Department of Laboratory Medicine and Pathobiology, University of Toronto; Peter Munk Cardiac Centre, University Health Network; Vector Institute, Toronto, Canada\\
Yao Zhang is with Shanghai AI Laboratory, 200232, Shanghai, China\\
Song Gu is with the Department of Image Reconstruction, Nanjing Anke Medical Technology Co., Ltd., 211113, Nanjing, China\\
Cheng Ge is with Ocean University of China, 266100, Qingdao, China\\
Shihao Ma and Adamo Young are with the Department of Computer Science, University of Toronto; Peter Munk Cardiac Centre, University Health Network; Vector Institute, Toronto, Canada\\
Cheng Zhu is with Tinavi Medical Technologies Co., Ltd., 100192, Beijing, China\\
Kangkang Meng is with University of Science and Technology Beijing, 100083, Beijing, China\\
Xin Yang is with the School of Biomedical Engineering, Health Science Center, Shenzhen University, 518055, Shenzhen, China\\
Ziyan Huang is with Shanghai AI Laboratory, 200232, Shanghai, China\\
Fan Zhang is with the Department of Radiological Algorithm, Fosun Aitrox Information Technology Co., Ltd., 200033, Shanghai, China\\
Wentao Liu is with the School of Artificial Intelligence, Beijing University of Posts and Telecommunications, 100876, Beijing, China\\
Yuanke Pan and Shoujin Huang are with Shenzhen Technology University, 518000, Shenzhen, China\\
Jiacheng Wang is with Xiamen University, 361005, Xiamen, China\\
Mingze Sun is with Alibaba, 100084, Beijing, China\\
Weixin Xu is with Infervision Medical Technology Co., Ltd., 100025, Beijing, China\\
Dengqiang Jia is with Hong Kong Centre for Cerebro-cardiovascular Health Engineering, 000000, Hong Kong, China\\
Jae Won Choi is with the Department of Radiology, Armed Forces Yangju Hospital, 11429, Yangju, Korea\\
Nat\'{a}lia Alves and Bram de Wilde are with the Department of Radiology, Radboudumc, 6525XZ, Nijmegen, Netherlands\\
Gregor Koehler is with the Department of Medical Image Computing, German Cancer Research Center, 69120, Heidelberg, Germany\\
% Haoran Lai is with Southern Medical University, 510515, Guangzhou, China\\
Yajun Wu is with ShenZhen Yorktal DMIT LLC, 518100, Shenzhen, China\\
Manuel Wiesenfarth is with the Division of Biostatistics, German Cancer Research Center (DKFZ), Heidelberg 69120, Germany\\
Qiongjie Zhu, Guoqiang Dong, and Jian He are with the Department of Nuclear Medicine, Nanjing Drum Tower Hospital, 210008, Nanjing, China\\
Junjun He is with Shanghai AI Laboratory, 200232, Shanghai, China\\
Hua Yang is with the Department of Radiological Algorithm, Fosun Aitrox Information Technology Co., Ltd., Shanghai, China\\
Huihua Yang is with the School of Artificial Intelligence, Beijing University of Posts and Telecommunications, Beijing, 100876, China\\
Bingding Huang is with the College of Big Data and Internet, Shenzhen Technology University, Shenzhen, 518188, China\\
Mengye Lyu is with the College of Health Science and Environmental Engineering, Shenzhen Technology University, Shenzhen, China\\
Yongkang Ma is with the Manteia Technologies Co.,Ltd, Xiamen, China\\
Heng Guo is with the Alibaba DAMO Academy, Beijing, China\\
Rongguo Zhang is with Infervision Medical Technology Co., Ltd., 100025, Beijing, China\\
Klaus Maier-Hein is with the Department of Medical Image Computing, German Cancer Research Center, 69120, Heidelberg, Germany\\
Bo Wang is with the  Peter Munk Cardiac Centre, University Health Network; Department of Laboratory Medicine and Pathobiology and Department of Computer Science, University of Toronto; Vector Institute, Toronto, Canada

\section*{Methods}\label{method}
\subsection*{Study design} The study design of the FLARE 2022 challenge was preregistered\cite{MA-FLARE22-Design} and passed peer review on the 25th International Conference on Medical Image
Computing and Computer-Assisted Intervention (MICCAI 2022). 
The dataset was retrospectively obtained from 53 different medical sources (Supplementary Table 1-3). During the development phase, participants could access the 50 labeled cases from one medical center and 2000 unlabeled cases from 21 medical centers. All the internal and external validation sets were hidden from participants. For fair comparisons, participants were not allowed to use additional data or pretrained models to develop their methods.

We launched the challenge on 15 March 2022 via the GrandChallenge platform. During the development phase, each team had three chances per day to make submissions to the online evaluation platform and get segmentation accuracy scores. Moreover, each team also had five chances to submit their algorithm docker containers to challenge organizers and obtain segmentation efficiency scores. During the validation phase, each team was required to submit the final algorithm by 15 July 2022. The algorithm should be packaged as a docker container and we independently evaluated the algorithm docker on the internal validation set that is blind to participants.

\subsection*{Data standardization}
All the images were de-identified and can be used in this challenge based on legal licenses (e.g., CC-BY, CC-BY-NC-SA). 
We normalized all the images to standard NIfTI format (\url{https://nifti.nimh.nih.gov/}) because it only contains necessary image information (intensity, orientation, spacing, origin) and without any patient metadata (e.g., age, gender). Compared to the well-known DICOM format (\url{https://www.dicomstandard.org/}), NIfTI aggregates multiple 2D slices to one 3D volume, which is easier to handle for developing AI algorithms. NIfTI has been the most popular data format in 3D medical image analysis challenges~\cite{MSD-Summary, bakas2018brats}.
The voxel orientation was standardized as canonical 'RAS', which means that the first, second, and third voxel axes go from left to Right, posterior to Anterior, and inferior to Superior, respectively.  

\subsection*{Annotation protocols}
The annotation process involved a team of five junior radiologists with one to five years of experience and two senior radiologists with more than 10 years of experience. The organs vary in different sizes, morphologies, and appearances. For example, the average liver volume ($1500 cm^3$) is around 300 times larger than the volume of the adrenal gland ($5 cm^3$). The esophagus, aorta, and inferior vena cava emerge as lumen structures, while the spleen and pancreas appear as irregular masses of tissues. Moreover, some organs are close to each other with low contrasts, such as the liver and stomach. Algorithms need to address all the difficulties at the same time since they occur simultaneously. 
Annotation quality is crucial to algorithm performance and it is necessary to reduce the annotation variability~\cite{Lena-Annotation} between annotators. Therefore, we provided detailed annotation instructions to the annotators and employed a hierarchical human-in-the-loop pipeline to enhance throughput and maintain consistent annotations.

The hierarchical annotation pipeline consisted of three stages to ensure accuracy and consistency throughout the annotation process. In the first stage, an annotation consensus was formulated by a senior radiologist, an oncologist, and a surgeon based on radiation therapy oncology group consensus (RTOG) panel guideline~\cite{goodman2012RTOG} and Netter's anatomical atlas~\cite{netter-atlas2014}.
Specifically, the liver contour should include all hepatic parenchyma and all liver lesions. The hepatic vessels inside the liver also need to be covered. If the vessels are located outside the liver (i.e., the entrance of portal hepatitis) based on the coronal view, they should be excluded from the liver contour. The kidney contour should include the renal parenchyma while excluding adjacent structures such as blood vessels and surrounding fat.
The spleen contour should include all splenic parenchyma and any splenic lesions. It should exclude adjacent structures such as the splenic vessels (arteries and veins), particularly those located outside the spleen. The pancreatic contour should encompass all pancreatic parenchyma including the head, body, and tail, as well as any pancreatic lesions. Exocrine, endocrine components, and the pancreatic duct need to be included, but the surrounding vessels and fat should be excluded. The aortic contour should include the entire lumen of the aorta, from the aortic root to the bifurcation. The aortic wall (including the aortic calcification) should also be included. The inferior vena cava contour should include the entire lumen and cover the walls. The adrenal gland contour should include the entire adrenal gland, both cortex and medulla, and any adrenal lesions. The gallbladder contour should encompass the entire gallbladder wall, including the body, fundus, and neck, as well as any gallstones or polyps. The cystic duct and the surrounding liver parenchyma should be excluded. The esophagus contour should include the entire esophageal wall, while adjacent structures such as the trachea, aorta, and surrounding fat and muscle should be excluded. The stomach contour should encompass the entire stomach wall including the fundus, body, antrum, and pylorus, as well as any gastric lesions. The duodenum contour should include the entire duodenal wall from the duodenal bulb to the ligament of Treitz, along with any duodenal lesions. It should exclude surrounding structures such as the head of the pancreas, common bile duct, and surrounding vasculature.
Before the annotation process, all annotators were required to learn and follow the annotation consensus. 

The second stage of our annotation pipeline involved a human-in-the-loop approach aimed at enhancing annotation throughput. To facilitate this process, we utilized five 3D U-Net models~\cite{isensee2020nnunet} trained via 5-fold cross-validation on existing abdomen CT datasets~\cite{Ma-2020-abdomenCT-1K, landman2015btcv}. Leveraging the predictions generated by these models, junior annotators performed manual refinements on 100 randomly selected predictions, which were then checked and revised by the senior radiologists.  
This process was iterated seven times until all the images were labeled by one of the junior annotators and further refined by the senior radiologists. 
In the external validation set (600 CT scans), the organ annotations of 512 CT scans were generated by us, and the remaining 84 CT scans were collected from recent public abdomen datasets~\cite{luo2022word-mia, amos-data, totalsegmentator}. Their annotations were used and manually refined to follow our annotation consensus. It is important to note that these publicly available annotations were accessible only after the challenge, thereby preventing participants from utilizing them during the challenge. 
In the third and final stage, we aimed to identify potential annotation errors. We trained a new set of U-Net models using five-fold cross-validation, with special attention given to images exhibiting low Dice Similarity Coefficient (DSC) scores ($\leq$0.75), which were then double-checked by the senior radiologists.

\subsection*{Evaluation platform}
All the submitted algorithms were evaluated on the same workstation. Specifically, the workstation is a Ubuntu 20.04 desktop with one central processing unit (CPU, Intel Xeon(R) W-2133 CPU, 3.60GHz × 12 cores), one graph processing unit (GPU, NVIDIA QUADRO RTX5000, 16G), 32G of memory, and 500G of hard disk drive storage. At the beginning of the challenge, we also released the versions of GPU Driver (510.60.02), CUDA (11.6), and Docker (20.10.13) to make sure that participants' algorithms are compatible with our evaluation platform.

\subsection*{Evaluation metrics}
We used five complementary metrics to quantitatively evaluate the segmentation accuracy, efficiency, and resource consumption of the algorithms (Supplementary).
Specifically, Dice Similarity Coefficient (DSC), the most popular segmentation metric~\cite{Lena2018rankings-NC}, was used to evaluate the spatial overlap between the segmentation mask and ground truth, which is defined by 
\begin{equation*}
    DSC(G, S) = \frac{2|G\cap S|}{|G| + |S|},
\end{equation*}
where $|\cdot|$ counts the number of voxels and $G$ and $S$ denote the ground truth and segmentation mask, respectively.
Normalized Surface Distance (NSD) was used to measure the boundary accuracy, which is defined by
\begin{equation*}
    NSD(G, S) = \frac{|\partial G\cap B_{\partial S}^{(\tau)}| + |\partial S\cap B_{\partial G}^{(\tau)}|}{|\partial G| + |\partial S|},
\end{equation*}
where $B_{\partial G}^{(\tau)} = \{x\in R^3 \, | \, \exists \tilde{x}\in \partial G,\, ||x-\tilde{x}||\leq \tau \}$ and $B_{\partial S}^{(\tau)} = \{x\in R^3 \,|\, \exists \tilde{x}\in \partial S,\, ||x-\tilde{x}||\leq \tau \}$  denote the border region of the ground truth and the segmentation surface at tolerance $\tau$, respectively. The tolerance $\tau$ is defined by measuring organ segmentation consistency between radiologists and revised by oncologists and surgeons based on their clinical requirements (Supplementary). We also found that our boundary tolerances are consistent with other independent inter-rater annotation variability studies~\cite{PediatricCT22Eval,MSD-Summary}. 
The running time was used to measure the segmentation efficiency which is defined by the duration $T$ between the algorithm docker container start and end.
In order to obtain precise metrics, the algorithms were evaluated one by one, and all the other applications on the workstation were closed during the running time.
The resource consumption was measured by GPU memory consumption and CPU utilization. Instead of only capturing the maximum consumption~\cite{FLARE21-MIA}, we measured all the cumulative resource consumption during the algorithm running time. In particular, we recorded the GPU memory and GPU utilization every 0.1s and defined two new metrics: Area under GPU memory-time curve (AUC\_GPU) and Area under CPU utilization-time curve (AUC\_CPU), which are defined by 
\begin{equation*}
    AUC\_GPU = \sum_{t=0}^{T} GPU_t
\end{equation*}
and 
\begin{equation*}
    AUC\_CPU = \sum_{t=0}^{T} CPU_t,
\end{equation*}
where $GPU_t$ and $CPU_t$ denote the GPU memory consumption and GPU utilization at timepoint $t$. 
If the algorithms get stuck, the corresponding metrics will be set to the worse values (DSC=0, NSD=0, Time=3600, AUC\_GPU=3600*(1024*10-2048)=29491200, AUC\_CPU=3600*100=360000). Among the 48 evaluated algorithms, only two algorithms got stuck during model inference because they did not optimize the model efficiency (e.g., upsampling the 3D CT scans to high resolution and directly loading the whole volume to memory rather than using a patch-based way).

\subsection*{Ranking scheme}
All metrics were used to compute the final ranking. We give GPU memory consumption a tolerance of 2048MB  because this kind of GPU is affordable ($\leq$100 USD) for most medical centers and personal computers. Thus, the GPU memory consumption at time $t$ was transformed by $\hat{GPU_t} = \max(0, GPU_t - 2048MB)$.
We assigned a half weight to CPU and GPU metrics, which can achieve a balance between segmentation accuracy and efficiency\&resource consumption in the ranking scheme. 
The internal validation set contains 200 cases. The corresponding ranking scheme includes three steps:
\begin{itemize}
    \item Step 1. Computing the five metrics (DSC, NSD, Running Time, AUC\_GPU, and AUC\_CPU) for each case.
    \item Step 2. Ranking the 48 algorithms for each case and each metric. Thus, each algorithm will have $200\times5$ rankings.
    \item Step 3. Computing the overall rank for each algorithm by averaging all the rankings (AUC\_GPU and AUC\_CPU metrics are weighted by 0.5).
\end{itemize}

\subsection*{Ranking stability and statistical analysis}
% https://www.nature.com/articles/s41598-021-82017-6
The challenge ranking should be independent of the specific datasets because it reflects the task performances of algorithms. We applied bootstrapping and computed Kendall’s $\tau$~\cite{kendall-tal} to quantitatively analyze the variability of our ranking scheme. Specifically, we first extracted 1000 bootstrap samples from the international validation set and computed the ranks again for each bootstrap sample. Then, the ranking agreement was quantified by Kendall’s $\tau$. 
Kendall’s $\tau$ computes the number of pairwise concordances and discordances between ranking lists. Its value ranges $[-1, 1]$ where -1 and 1 denote inverted and identical order, respectively. A stable ranking scheme should have a high Kendall’s $\tau$ value that is close to 1.

To compare the performance of different algorithms, we performed Wilcoxon signed rank test because it is a paired comparison. Results were considered statistically significant if the $p-$value is less than 0.05.

The above analysis was performed with ChallengeR~\cite{challengeR}, Python 3~\cite{python}, Numpy~\cite{harris2020numpy}, Pandas~\cite{reback2020pandas}, Scipy~\cite{virtanen2020scipy}, PyTorch~\cite{paszke2019pytorch}, and matplotlib~\cite{matplotlib}.

\subsection*{Data availability}
The full development set has been released on the challenge website \url{https://flare22.grand-challenge.org/} and participants can access the data by registering on the Grand Challenge website. The hidden internal validation set and external validation sets will be publicly available on the challenge website after peer review. 

\subsection*{Code availability}
The code, method description, and docker containers of the top teams are available at \url{https://flare22.grand-challenge.org/awards/}.
The evaluation code is available at \url{https://github.com/JunMa11/FLARE}.

\bibliographystyle{IEEEtran}
\bibliography{JunRef}

\end{document}